\documentclass[letterpaper, 10 pt, conference]{ieeeconf}

\makeatletter
\def\munderbar#1{\underline{\sbox\tw@{$#1$}\dp\tw@\z@\box\tw@}}
\makeatother

\usepackage{tabularx}
\usepackage{algpseudocode}
\usepackage{algorithm}
\usepackage{algorithmicx}
\usepackage{lipsum} 
\usepackage{arydshln}
\usepackage[dvips]{color}
\usepackage{comment}
\usepackage{todonotes}
\usepackage{epsf}
\usepackage{epsfig}
\usepackage{times}
\usepackage{epsfig}
\usepackage{graphicx}
\usepackage{bbold}
\usepackage{mathrsfs}
\usepackage{amssymb}
\usepackage{epstopdf}
\newfloat{algorithm}{t}{lop}
\usepackage{amsmath}

\usepackage[symbol]{footmisc}

\usepackage{hyperref}
\usepackage{cite}
\usepackage{dsfont}

\usepackage{amsmath,epsfig,amssymb,algorithm,algpseudocode,amsthm,cite,url}
\usepackage{subcaption}
\allowdisplaybreaks
\usepackage{csquotes}
\usepackage{verbatim}
\usepackage[english]{babel}
\usepackage{amsmath,amssymb}

\usepackage{hyperref}
\usepackage[justification=centering]{caption}
\usepackage{verbatim}

\begin{document}
	\IEEEoverridecommandlockouts
	\title{ BLEBeacon:  A Real-Subject Trial Dataset from Mobile BLE-beacons}
		\title{ BLEBeacon:  A Real-Subject Trial Dataset from Mobile \\Bluetooth Low Energy Beacons}

	\author{Dimitrios Sikeridis$^{1}$, Ioannis Papapanagiotou$^{2,1}$, and Michael Devetsikiotis$^{1}$\\	{\tt\small \{dsike, ipapapa, mdevets\}@unm.edu}
		\thanks{$^{1}$Department of Electrical and Computer Engineering, The University of New Mexico, 
Albuquerque, NM, USA
			}%
		\thanks{$^{2}$Platform Engineering, Netflix, Los Gatos, CA, USA}
	\vspace*{-100cm}}
\makeatletter
\patchcmd{\@maketitle}
{\addvspace{0.5\baselineskip}\egroup}
{}
{}
\makeatother
	\maketitle
	\thispagestyle{empty}
	\pagestyle{empty}

	\IEEEpeerreviewmaketitle
\begin{abstract}
The BLEBeacon dataset\footnote[1]{The BLEBeacon dataset can be found online in: \url{https://github.com/dimisik/BLEBeacon-Dataset}, and as part of the CRAWDAD Community Resource repository \cite{unm-blebeacon-20190312}.} is a collection of Bluetooth Low Energy (BLE)  advertisement packets/traces generated from BLE beacons carried by people following their daily routine inside a university building. A network of Raspberry Pi 3 (RPi)-based edge devices were deployed inside a multi-floor facility continuously gathering BLE advertisement packets and storing them in a cloud-based environment \cite{sikeridis2017comparative}. The data were collected during an IRB (Institutional Review Board forhe Protection  of  Human Subjects in Research) approved  one-month trial. Each facility occupant/participant was handed a BLE beacon to carry with him at all times. The focus is on presenting a real-life realization of a location-aware sensing infrastructure, that can provide insights for smart sensing platforms, crowd-based applications, building management, and user-localization frameworks. This work describes and documents the published BLEBeacon dataset.
\end{abstract}
\section{Introduction}

Advanced smart infrastructures enable creative ways to enhance life quality and provide information flow towards facility owners, supporting occupant localization services, detection of human patterns, and enabling facility-occupant interaction applications \cite{inaya2017real, sikeridis2017occupant}. We describe a dataset of Bluetooth Low Energy (BLE) advertisement readings \cite{heydon2013bluetooth} collected during an IRB-approved one-month trial with real participants. The central component in our approach is a moving BLE beacon architecture that deploys static scanners to intercept advertisement packets. Gathered packets supported two functionalities (see Section \ref{sec:3}) and were forwarded to a central server to be used for real-time visualizations, or occupant activity recognition.

Real-subject-based datasets are valuable for a wide spectrum of research fields. The majority of existing studies on human activity sensing focus either on restrained settings (small scale) or is employed for a small duration (a couple of days) \cite{cabrero2017cwi}. This work provides a dataset extracted from a large-scale multi-floor setting and for an extensive one-month long experiment period. The following sections describe the experimental setup, the real-subject trial, and the dataset contents.

\section{Bluetooth Low Energy Beacons}

The BLE standard was proposed by the Bluetooth Special Interest Group (SIG) and is designed for low transmission power, short data burst communications. The protocol is operating at the 2.4 GHz frequency band with 40 channels of 2MHz spacing (3 for advertising, 37 for data transmissions). While BLE supports classic master/slave connection paradigm its extremely energy efficient nature drew attention to a new class of mini-scale devices the BLE beacons. These devices operate by periodically broadcasting packets/identifiers at a specific time interval and transmission power.
On the receiving end of those transmissions there can be BLE enabled devices able to utilize information such as the Received Signal Strength Indicator (RSSI) to support various application with Micro-Location, geofencing \cite{zafari2016microlocation}, and occupant tracking \cite{sikeridis2017occupant} being some cases in point. The sustainability of such use case is further supported by the low power consumption of BLE beacons that allows the devices to be powered by a coin cell battery for years before any need for reconfiguration.

In light of the above, initially, Apple with iBeacon \cite{iBeaconProtocol} and later Google with Eddystone \cite{eddy} standardized BLE beacon protocols (essentially the advertising packet formats). A BLE beacon that utilizes the iBeacon profile transmits a message that contains three pieces of data:
(a) A Universally Unique Identifier (UUID) which is the identity of the beacon (b) a Major value that denotes general spatial information, and (c) a Minor value that denotes more specific spacial information.
Regarding Eddystone the operation is similar to the iBeacon with the exception that it can support four different payload types in its frame format: (a) Unique ID (UID) frame - denotes identity and consists of two parts Namespace (10 bytes), and Instance (6 bytes) (b) URL address frame (URL) - carries a URL that the BLE end device can directly open (c) Sensor Telemetry frame (TML) - can be used to send sensor data, and (d) Ephemeral Identifier frame - the beacon broadcasts a continuously changing identifier that can be resolved to data by an end BLE device that shares a specific key with the beacon (security feature). More information regarding BLE Beacons and their applications in Internet of Things settings can be found in \cite{zafari2016microlocation , zafari2017ibeacon, jeon2018ble}.

\begin{figure}[t]
\centering
\includegraphics[width=\columnwidth]{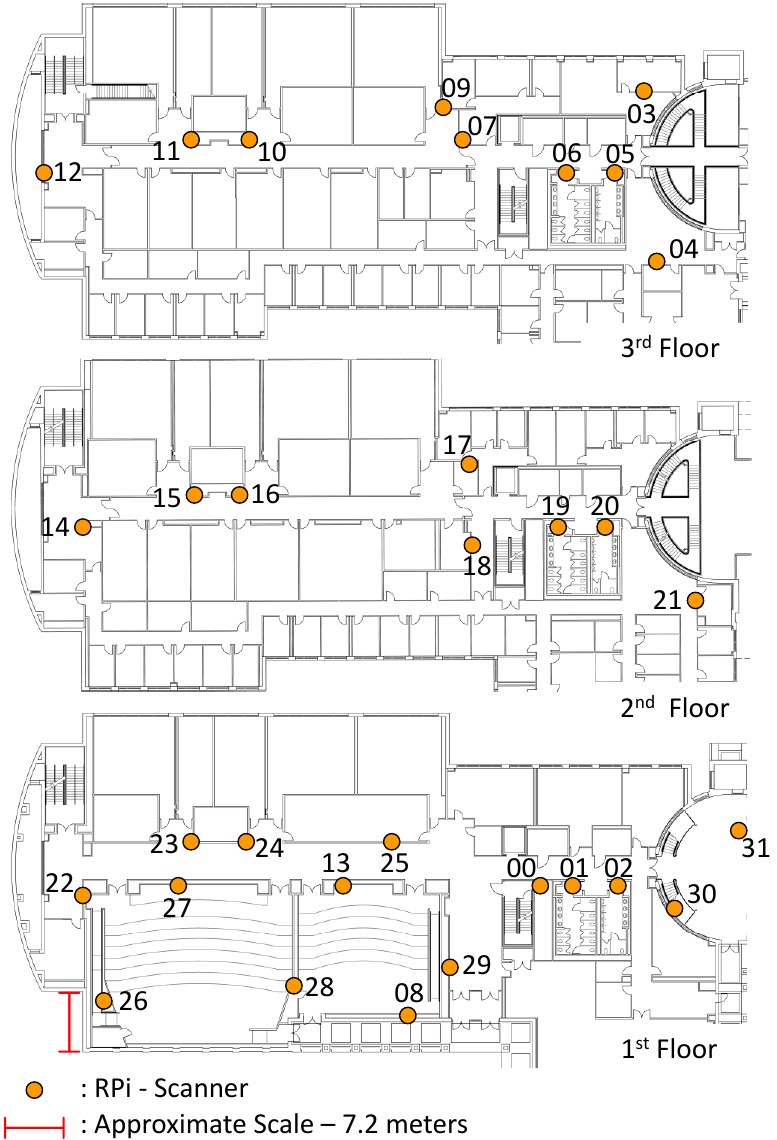}
\caption{Location of RPis - Facility topology.}
\label{fig:loc}
\end{figure}

\section{Sensing Infrastructure and Trial}

In this section we document the utilized devices and elaborate on the deployed infrastructure that collects the trial data.

\subsection{BLE Advertising Devices}

Facility occupants carry off-the-shelf BLE-based beacons that continuously transmit BLE advertisement packets. We used the Gimbal Series 10 iBeacon \cite{gimbal, iBeaconProtocol} configured to broadcast under the same 16-byte universally unique identifier (UUID). Each beacon is distinguished by a 4-byte identifier inside the manufacturer BLE advertisement packets. The periodic transmission rate for each beacon is set to 1 Hz, with omni-directional antenna propagation setting, and transmission power of 0 dBm.

\subsection{Sensing Infrastructure}

The deployment of the sensing infrastructure aimed for easy installation and cost efficiency. Thus, the backbone of the system is the Raspberry Pi 3 (RPi), which is able to listen to the BLE advertisement channels, and collect all generated packets. All utilized RPi edge devices are connected through WiFi and act as MQTT (Message Queuing Telemetry Transport) \cite{mqttProtocol} clients transmitting information to an MQTT broker hosted by the server. The server performs data management, validating and storing information in a MariaDB SQL database. 

Thirty two RPis were installed in three floors within NCSU Centennial Campus - Engineering Building II in order to support the experiment. The exact location of the RPis in the three-floor setting is shown in Fig. \ref{fig:loc}.

\begin{figure}[t]
\centering
\includegraphics[width=\columnwidth]{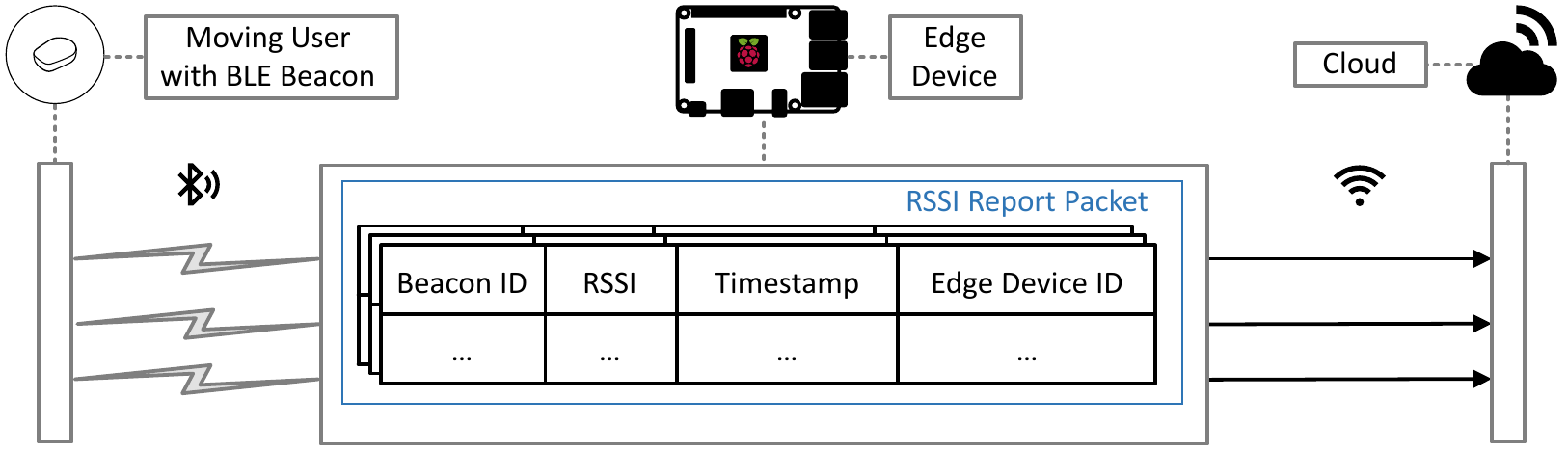}
\caption{RSSI report operation.}
\label{fig:ARCH}
\end{figure}

\begin{figure}[t]
\centering
\includegraphics[width=\columnwidth]{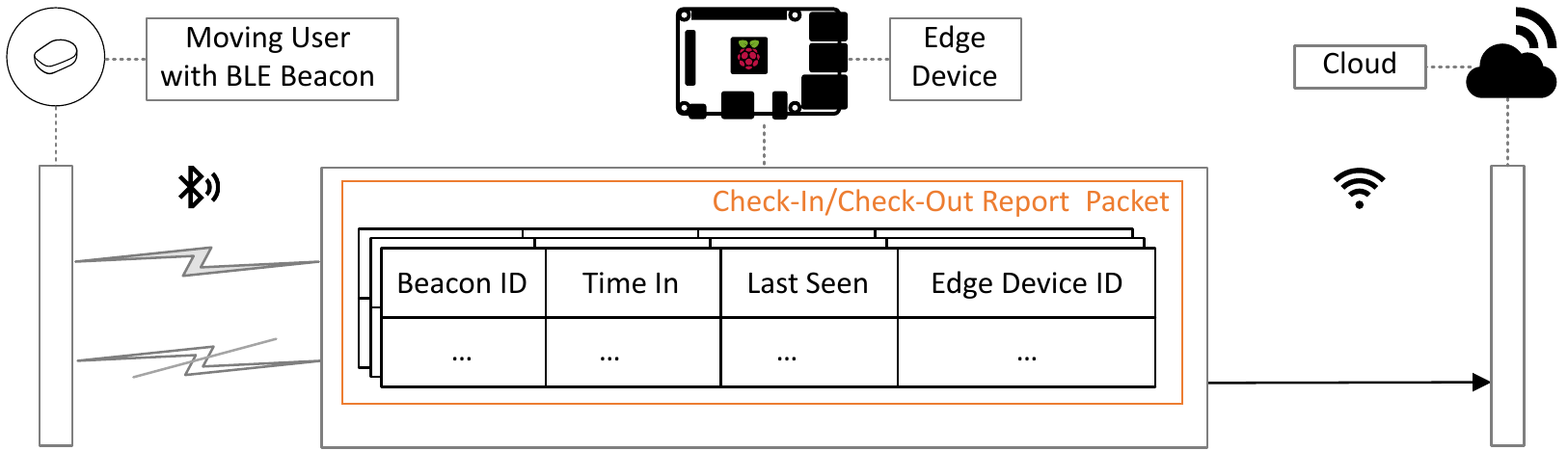}
\caption{ Check-In/Check-Out report operation.}
\label{fig:check}
\end{figure}

\begin{table}[!h]
\scriptsize
\renewcommand{\arraystretch}{1.3}
\caption{Trial \& Sensing Infrastructure Information}
\label{Table:components}
\centering
\begin{tabular}{||l||l||}
\hline 
\bf{BLE Beacon Type}               & Gimbal Beacon Series 10                  \\ \hline 
\bf{Number of Beacons/Participants}  & 46 \\ \hline
\bf{BLE Beacon Scanner Type}       & Raspberry Pi 3    \\ \hline
\bf{Number of Scanners}       & 32   \\ \hline
\bf{Number of Floors}       & 3    \\ \hline
\bf{Trial Dates}       & 09/15/2016 - 10/17/2016    \\ \hline
\end{tabular}
\end{table}

\subsection{Operation}

Regarding system operation two approaches were utilized in parallel:
\begin{enumerate}
    \item {\it RSSI Report}: all advertisement packet receptions from occupant beacon devices are directly reported to the server with a message that contains the beacon/user ID, the packet's Received Signal Strength Indicator (RSSI), a reception timestamp, and finally the ID of the RPi that received the advertisement, as seen in Fig. \ref{fig:ARCH}.
    \item {\it Check-In/Check-Out Report}: each RPi scanner continuously manages a list of current occupants/users in its proximity. A {\it check in} timestamp is created during occupant's initial entry, and while this beacon is still being detected by the RPi, a {\it last seen} timestamp is updated. When the beacon is no longer detected (advertisement packets are no longer being received) a Check-In/Check-Out report packet is created and sent to the server containing the beacon/user ID, the {\it check in} timestamp, the {\it last seen} timestamp, and finally the ID of the RPi as seen in Fig. \ref{fig:check}. A thirty second period is used to ensure that the occupant exited the RPi proximity.
\end{enumerate}

\subsection{Real-Subject Trial}

Following the system architecture described above, an IRB-approved trial with 46 participants took place from September 15 to October 17 of 2016. Participants included frequent occupants of the building that carried a BLE beacon with them at all times during their usual routines. The experiment considered all three-floors (see Fig. \ref{fig:ARCH}) and the core idea was to get insights on repeated occupant behavior and patterns in relation to the facility environment.
Table \ref{Table:components} summarizes basic trial and sensing infrastructure information.

\section{BLEBeacon Dataset}\label{sec:3}

The discussed dataset consists of two files, one containing the trial readings from the RSSI report operation ({\it RSSI Report} file) and the other from the Check-In/Check-Out report operation 
({\it Check-In Check-Out Report} file). No participant personal information were kept or made available to retain personal privacy. The {\it RSSI Report} file contains the following entries:
\begin{itemize}
    \item {\it Entry\_id}: unique identifier of a packet in the dataset.
    \item {\it Beacon\_id}: unique identifier of the occupant/beacon.
    \item {\it RSSI}: the Received Signal Strength Indicator (RSSI) in dB.
    \item {\it Timestamp}: Date (Month/Day/Year) and Unix time (Hour:Second) of the advertisement packet reception moment from the Rpi.
    \item {\it RPi\_id}: RPi that received the packet (see Fig. \ref{fig:ARCH}).
\end{itemize}
The {\it Check-In/Check-Out Report} file contains {\it Entry\_id}, {\it Beacon\_id}, and  {\it RPi\_id} as described above with the addition of two entries namely:
\begin{itemize}
    \item {\it In\_time}: Date (Month/Day/Year) and Unix time (Hour:Second) of the moment a user enters the RPi's vicinity and the first advertisement packet is received.
    \item {\it Out\_time}: Date (Month/Day/Year) and Unix time (Hour:Second) of the last advertisement packet received from the same user by the specific RPi.
\end{itemize}

Due to the architecture of the sensing infrastructure where several RPi scanners were deployed in close proximity, a single BLE advertisement packet from an occupant/beacon can be received fro multiple RPis. This creates multiple entries of the same packet in the database. Such packets are timestamped during the reception moment at the RPi scanner. The RSSI measurements from different RPis whose locations are known can be used to identify a user's exact location inside the facility. An preliminary analysis of the dataset is described in \cite{inaya2017real}. The BLEBeacon dataset is a part of the CRAWDAD repository \cite{unm-blebeacon-20190312}. Work related to the BLEBeacon dataset can be found in \cite{inaya2017real, sikeridis2017cloud, sikeridis2017occupant, sikeridis2018unsupervised}.

\section{Conclusion}

This report accompanies and documents the BLEBeacon Dataset, a collection of BLE advertisement packets gathered from a three-floor sensing infrastructure accommodating real-participants carrying iBeacons, following their routines during a one-month period. Possible uses of the dataset include network behavior and reliability detection in similar sensing environments, user mobility pattern extraction due to the experiment's length, occupancy clustering with group identification/monitoring, and provision of facility management or crowd monitoring application insights considering real-life conditions.

\section*{Acknowledgment}
 \vspace{-1mm}
\def\baselinestretch{0.85}

This research was supported by an IBM Faculty Award.
The authors are grateful to Mahdi Inaya and Michael Meli for carrying out the system deployment and installation, the NC State ECE Department for allowing us to host the trial in the EBII building, and to the gracious ECE trial participants.





\bibliographystyle{IEEEtran}
\bibliography{IEEEabrv,references}
\end{document}